\documentclass[aps,pre,reprint,10pt,superscriptaddress,showpacs]{revtex4-1}
\usepackage{amsfonts}
\usepackage{amsmath}
\usepackage{amssymb}
\usepackage{graphicx}%\[  \]
\usepackage{subfigure}
\usepackage{color}
\begin{document}
\title{Multidimensional ion-acoustic solitary waves and shocks   in quantum plasmas}
\author{A. P. Misra}
\email{apmisra@visva-bharati.ac.in, apmisra@gmail.com}
\affiliation{Department of Mathematics, Siksha Bhavana, Visva-Bharati University, Santiniketan-731 235, West Bengal, India}
\author{Biswajit Sahu}
 \email{biswajit$_$sahu@yahoo.co.in}
\affiliation{Department of Mathematics, West Bengal State University, Barasat, Kolkata-700 126, India}
\begin{abstract}
The nonlinear theory of two-dimensional ion-acoustic (IA) solitary waves  and shocks (SWS) is revisited  in a dissipative quantum plasma. The  effects of dispersion, caused by the charge separation of electrons and ions  and the quantum force associated with the Bohm potential for degenerate electrons, as well as, the dissipation    due to the ion kinematic viscosity are considered.   Using the  reductive perturbation technique,   a  Kadomtsev-Petviashvili Burgers (KPB)-type equation, which governs  the evolution of small-amplitude SWS in quantum plasmas, is derived, and its different solutions are obtained and analyzed.  It is shown that the KPB equation can admit either compressive or rarefactive SWS according to when $H\lessgtr2/3$, or the particle number density satisfies $n_0\gtrless 1.3\times10^{31}$ cm$^{-3}$, where $H$ is the ratio of the electron plasmon energy to the Fermi energy densities.    Furthermore, the properties of large-amplitude stationary shocks are studied  numerically in the   case when the wave dispersion due to charge separation is negligible. It is also shown that a transition from monotonic to oscillatory shocks occurs by the effects of the quantum parameter $H$.
\end{abstract}
\maketitle

%%%
%%%%%%%%%%%
\section{Introduction}
Over the last decade, there has been much attention in investigating the nonlinear electrostatic and electromagnetic structures in quantum plasmas. The latter were first studied by Pines \cite{pine} in regimes of  high-densities and   low-temperatures   as compared to classical plasmas. In these regimes of  quantum plasmas, the thermal de Broglie wavelength of the charge carriers becomes comparable to the dimension  of the system, and so the quantum statistical (Fermi-Dirac pressure) as well as mechanical (such as tunneling associated with the Bohm potential) effects   must be taken into account in the description of dynamics of charged particles such as electrons, positrons or holes. Recently, the field of quantum plasmas, where the dominated wave nature of electrons gives rise to collective effects, has become an intense field of investigation, having applications in microelectronics \cite{q1},   quantum dots and quantum wires \cite{q2},   quantum wells,   carbon nanotubes and quantum diodes \cite{q3},   nonlinear optics \cite{q4}, as well as,  in laser plasma experiments \cite{q5,q6,q7}.  On the other hand, quantum plasmas are of great importance in dense astrophysical environments such as those in white dwarf stars and magnetars \cite{wh1} etc. Moreover, it has recently been experimentally shown that quantum effects are  important in inertial confinement fusion (ICF) plasmas \cite{hora,conf}. There has also been   a great majority of recent activities in the field of quantum hydrodynamics (QHD) and quantum magnetohydrodynamics (QMHD) \cite{mhd1,mhd2,mhd3,mhd4,mhd5,mhd6,mhd7}, which highlight the unexpected nonlinear wave features in quantum plasmas. The QHD model generalizes the fluid model with the inclusion of quantum statistical pressure and quantum diffraction (associated with the Bohm potential) terms. The validity of QHD model is limited to those systems that are large compared to the Fermi Debye lengths of the species in the system.
 
Investigations on electrostatic nonlinear waves, especially, solitary waves and shocks (SWS)   in plasmas  have attracted more attention because of their vital roles in research in space, Earth, as well as, in laboratory environments \cite{sok1,sok2}. Nonlinear structures such as SWS in plasmas have been observed in the laboratory, and also in condensed matter plasmas \cite{con} and colloidal suspension \cite{col}. When a medium has both dispersive and dissipative properties, the nonlinear propagation of electrostatic waves is governed by a Korteweg-de Vries Burgers (KdVB) equation in one dimension  and by a Kadomtsev Petviashvili Burgers (KPB) equation in   two-dimensional space. The dissipative Burgers' term in both the nonlinear KdVB and KPB equations arises  due to,  e. g., the effect of  kinematic viscosity  among the plasma constituents. However, when the dissipation overwhelms the dispersion, and when the dissipative effect is in balance with the nonlinearity, we indeed have the possibility of shock waves. In the absence of
dissipation (or if the dissipation is weak at the characteristic dynamical time scales of the system), the balance between nonlinear and dispersive effects can result into the formation of  solitons.
 
A number of theoretical investigations  have been carried out for understanding the collective processes as well as the formation of nonlinear coherent structures not only in classical plasmas (See, e.g., Refs. \citep{c1,c2,c3,c4}), but also  in dense quantum plasmas \cite{th1,th2,th3,th4,th5,th6,th7,th8,th9,misra2009}. For example, Mahmood \textit{et al.} \cite{mah} investigated the ion-acoustic (IA) solitary waves in quantum electron-ion (e-i)   plasmas by employing Sagdeev potential approach. Masood \cite{mas} studied the linear and nonlinear propagation of obliquely propagating magnetoacoustic waves in a dissipative quantum e-i  magnetoplasma using a QMHD model. Recently, Akhtar and Hussain \cite{akh} investigated quantum IA shock waves in a negative ion quantum plasma on the basis of a QHD model. More recently, the behavior of quantum dust ion-acoustic (DIA) shocks in a plasma including inertialess quantum electrons and positrons, classical cold ions and stationary negative dust grains are investigated by Rouhani \textit{et al.} \cite{rou}.  However,  all these investigations are limited to one-dimensional (1D) planar geometry which may not be a realistic situation in laboratory devices, since the waves observed in laboratories are certainly not bounded in one-dimension. 

The purpose of this present work is to study the nonlinear propagation of IA solitary and shock waves in dissipative  quantum  e-i plasma in a two-dimensional planar geometry. Both the small- and large amplitude electrostatic perturbations are considered. We show that these perturbations can develop into compressive or rarefactive SWS depending on whether the dispersive effects caused by the density correlation due to quantum fluctuations as well as by the charge separation of particles (deviation from quasineutrality)  is stronger or weaker than the effects of dissipation associated with the ion kinematic viscosity.     
%%%%%
\section{Theoretical model and derivation of KPB equation }
We consider the nonlinear propagation of ion-acoustic solitary waves   and shocks  in a  quantum plasma consisting of  degenerate  electrons providing all the pressure (restoring force) but none of the mass (inertialess) and positively charged inertial ions providing almost all the mass and none of the pressure. Any speed involved in the
plasma flow is assumed to be much lower than the quantum ion-acoustic (QIA) speed. At equilibrium, both electrons and ions have equal number density, say $n_0$. Thus, in  a collisionless
and unmagnetized quantum plasma the basic   set of quantum hydrodynamic equations  are \cite{hass2011}
%%%%%%%

\begin{equation}
\frac{\partial n_i}{\partial t}+\nabla\cdot(n_i{\bf v})=0,  \label{cont-eq}
\end{equation}

\begin{equation}
\frac{d{\bf v}}{dt}=-\frac{e}{m_i}\nabla \phi+\frac{\mu}{m_in_i}\nabla^2{\bf v},  \label{moment-ion}
\end{equation}

\begin{equation}
0=\frac{e}{m_e}\nabla \phi  -\frac{\nabla p_e }{m_en_e} +\frac{\hbar^2}{2m_e}%
\nabla\left(\frac{\nabla^2\sqrt{n_e}}{\sqrt{n_e}}\right),  \label{moment-elec}
\end{equation}

\begin{equation}
\nabla^2 \phi =\frac{e}{\epsilon_0}\left(n_{e}-n_{i}\right),  \label{poiss-eq}
\end{equation}%
which can be written in nondimensional form in two space dimensions as 
%%%%%%%%%
\begin{equation}
\frac{\partial n_i}{\partial t}+\frac{\partial }{\partial x}(n_iu)+\frac{%
\partial }{\partial y}(n_iv)=0,  \label{cont-eq-ni}
\end{equation}
\begin{equation}
\frac{du}{dt}=-\frac{\partial \phi }{\partial x}+\eta\triangle u,  \label{moment-ion-u}
\end{equation}
\begin{equation}
\frac{dv}{dt}=-\frac{\partial \phi }{\partial y}+\eta\triangle v,  \label{moment-ion-v}
\end{equation}
\begin{equation}
0=\phi +\frac{1}{2}\left( 1-n_{e}^{2/3}\right) +\frac{H^{2}}{2\sqrt{n_{e}}}%
\triangle \sqrt{n_{e}},  \label{moment-electron}
\end{equation}%
\begin{equation}
\triangle \phi =n_{e}-n_{i}.  \label{poisson-eq}
\end{equation}%
Here, $d/dt\equiv \partial/\partial t+u \partial/\partial x+v\partial/\partial y$ is the total derivative, $\triangle\equiv\partial^2/\partial x^2+\partial^2/\partial y^2$ is the Laplace operator,  $n_{e(i)}$ is the electron (ion) number density normalized by   $n_{0}$, ${\bf v}\equiv\left(u,~v\right)$ is the ion velocity  normalized by the QIA speed $c_{s}=\sqrt{k_{B}T_{Fe}/m_{i}}$ with $k_{B}$ denoting the Boltzmann constant, $m_{i}$  the ion mass, $T_{Fe}\equiv \hbar ^{2}(3\pi ^{2}n_{0})^{2/3}/2k_{B}m_{e}$ the electron Fermi temperature  and $\hbar$  the reduced Planck's constant. Also,   $\mu$ is the constant coefficient of the dynamical (shear) viscosity such that $\mu/m_in_0$ is that of the ion kinematic viscosity and $\eta=\mu\omega_{pi}/k_BT_{Fe}n_0$ is the nondimensional viscosity parameter which typically depends on the number density $n_0$. Also, $H=\hbar \omega _{pe}/k_{B}T_{Fe}$ is  the ratio of
the `plasmon energy density' to the Fermi thermal energy in which $\omega
_{pj}=\sqrt{n_{0}e^{2}/\varepsilon _{0}m_{j}}$ is the plasma oscillation frequency for
the $j$-th species particle. Furthermore, $\phi$ is the electrostatic potential
normalized by $k_{B}T_{Fe}/e$. The space and time variables are, respectively,
normalized by the Debye length  $\lambda_{Fe}\equiv c_{s}/\omega _{pi}$ and the ion plasma period $\omega^{-1} _{pi}$.  Note that Eq. \eqref{moment-electron} is obtained after integrating   the   momentum balance equation \eqref{moment-elec} for degenerate electrons and using the boundary conditions: $\phi\rightarrow0,~n_e\rightarrow1$ as $x,~y\rightarrow\pm\infty$. In this equation we have considered the pressure gradient as $\nabla p_{e}$, where  $p_e$ is  pressure given by the following equation of state for nonrelativistic  degenerate electrons \cite{chandrasekhar1935,misra2012}
\begin{equation}
p_{e}=\frac{1}{5}\frac{m_{e}V_{Fe}^{2}}{n_{0}^{2/3}}n_{e}^{5/3},  \label{pressure-electron}
\end{equation}%
where $V_{Fe}\equiv \sqrt{k_{B}T_{Fe}/m_{e}}$ is the Fermi thermal speed of
electrons. Furthermore, the term proportional to $H$ in Eq. \eqref{moment-electron} appears due to the quantum force ${\bf F}_q\equiv -\nabla V_B$ associated with the Bohm potential (tunneling effect)   given by 
\begin{equation}
V_{B}=-\frac{\hbar^2}{2m_e}\frac{1}{\sqrt{n_e}}\triangle\sqrt{n_e}. \label{bohm-potential}
\end{equation} 
Now, in the small-amplitude limit,   Fourier analyzing the linearized  basic equations \eqref{cont-eq-ni}-\eqref{poisson-eq} we obtain the following dispersion law for QIA waves
\begin{equation}
\left(\omega+i\eta\right)^2=\frac{k^2\Lambda}{1+k^2\Lambda}, \label{disp-relation}
\end{equation}
where $\omega$ and $k$ are the wave frequency and wave number of perturbations (oscillations), and $\Lambda=1/3+H^2k^2/4$. Equation \eqref{disp-relation} shows that the dispersion curves get modified by the effects of the quantum parameter $H$. The  wave becomes unstable due to the finite effect of the ion  viscosity $\eta$. However, in   the long-wavelength limit ($k\rightarrow0$) and in absence of the viscosity effect,  the phase speed of the wave becomes $\lambda\equiv\omega/k=1/\sqrt{3}$, i.e.,  constant. In other words,   QIA waves become  dispersionless when the system   length scale is much larger than the Fermi Debye length $\lambda_{Fe}$.  This  implies that in a frame moving with the speed $\lambda$, the time derivatives of all physical quantities should vanish. Thus, for a  finite $\epsilon$ with $0<\epsilon\lesssim1$, one can observe slow variations of the wave amplitude in the moving frame of reference. So,  we introduce the new variables for the space and the time   as \cite{taniuti1969,mas}
\begin{equation}
\xi =\epsilon ^{1/2}(x-\lambda  t),~\zeta=\epsilon y,~\tau =\epsilon^{3/2}t,
\label{stretch-coord}
\end{equation}%
where the phase speed $\lambda$ (normalized by $c_{s}$) equals $1/\sqrt{3}$.   The dynamical variables are expanded as \cite{taniuti1969,mas}
\begin{equation}
\begin{split}
n_{j}=&1+\epsilon n_{j}^{(1)}+\epsilon ^{2}n_{j}^{(2)}+\cdots, \\
u=&\epsilon u^{(1)}+\epsilon ^{2}u^{(2)}+\cdots,   \\
v=&\epsilon ^{3/2}v^{(1)}+\epsilon ^{5/2}v^{(2)}+\cdots,  \\
\phi=&\epsilon \phi ^{(1)}+\epsilon ^{2}\phi ^{(2)}+\cdots,
\label{expansion}
\end{split}
\end{equation}
where we have considered the   perturbations for the transverse  velocity component $v$ as higher-order effects, i.e., weaker than those of the longitudinal component $u$. This is due to the fact that since the wave propagation is assumed to propagate along the $x$-axis or $\xi$ direction  in the moving frame of reference, the effects of  dispersion due to separation of electron and ion charges and the quantum tunneling associated with the Bohm potential on QIA waves will appear   only in the $\xi$ direction, i.e., along the direction of the ion velocity component $u$.   We also assume that the effect  of the   viscosity is small and the constant $\eta$ is the same for both the velocity components $u$ and $v$, i.e.,  $\eta =\epsilon ^{1/2}\eta_{0}$, where $\eta _{0}$ is  of the order of unity.

Next,  we  substitute the expansions from Eq. \eqref{expansion}  into the basic nondimensional equations \eqref{cont-eq-ni}-\eqref{poisson-eq}  and equate the terms in different powers of $\epsilon$. In the lowest-order of $\epsilon$, we obtain the following expressions for the first-order quantities  
\begin{equation}
 n_{e}^{(1)}=n_{i}^{(1)}=3\phi^{(1)},~u^{(1)}=v^{(1)}=\sqrt{3}\phi^{(1)}, \label{first-order}
\end{equation} 
together with $\lambda=1/\sqrt{3}$, already considered in the stretching \eqref{stretch-coord}.
From the next order of $\epsilon $, we obtain the following equations 
\begin{equation}
-M\frac{\partial n_{i}^{(2)}}{\partial \xi }+\frac{\partial n_{i}^{(1)}}{%
\partial \tau }+\frac{\partial u^{(2)}}{\partial \xi }+\frac{\partial
(u^{(1)}n_{i}^{(1)})}{\partial \xi }+\frac{\partial v^{(1)}}{\partial \zeta }%
=0, \label{2-order1}\end{equation}
\begin{equation}
-M\frac{\partial u^{(2)}}{\partial \xi }+\frac{\partial u^{(1)}}{\partial
\tau }+u^{(1)}\frac{\partial u^{(1)}}{\partial \xi }=-\frac{\partial \phi
^{(2)}}{\partial \xi }+\eta _{0}\frac{\partial ^{2}u^{(1)}}{\partial \xi ^{2}%
}, \label{2-order2}\end{equation}
\begin{equation}
\phi ^{(2)}+\frac{H^{2}}{4}\frac{\partial ^{2}n_{e}^{(1)}}{\partial \xi ^{2}}%
-\frac{1}{3}n_{e}^{(2)}+\frac{1}{18}\left(n_{e}^{(1)}\right)^2=0,  \label{2-order3} 
\end{equation}
\begin{equation}
\frac{\partial ^{2}\phi ^{(1)}}{\partial \xi ^{2}}=n_{e}^{(2)}-n_{i}^{(2)} \label{2-order4}.
\end{equation}
Finally, eliminating the second-order quantities from Eqs. \eqref{2-order1}-\eqref{2-order4},  and substituting the expressions for the first-order quantities from Eq. \eqref{first-order} into the resulting equation, we obtain  the following  KPB equation 
\begin{equation}
\frac{\partial }{\partial \xi }\left(\frac{\partial \phi }{\partial \tau }+A\phi 
\frac{\partial \phi }{\partial \xi }+B\frac{\partial ^{3}\phi }{\partial \xi
^{3}}-C\frac{\partial ^{2}\phi }{\partial \xi ^{2}}\right)+D\frac{\partial
^{2}\phi }{\partial \zeta ^{2}}=0, \label{kpb-eq} 
\end{equation}
where $\phi\equiv \phi^{(1)}$, $A=4/\sqrt{3}$ is the nonlinear coefficient, which appears due to finite-amplitude effects of first-order perturbed quantities normalized by their equilibrium values and $B=\sqrt{3}\left(4/9-H^2\right)/8$ is the coefficient of dispersion arising due to the deviation from quasineutrality and the quantum tunneling effects. Furthermore, the dissipative coefficient $C=\eta_0/2$ appears due to the ion  viscosity and  $D=1/2\sqrt{3}$ is the effect of weak transverse perturbations (i.e., along the $\zeta$-axis). Equation \eqref{kpb-eq} describes the evolution of small-amplitude   ion-acoustic SWS in quantum plasmas. 
\section{Solution of KPB equation,   its stability and numerical results}
\subsection{Solution}
Here, we obtain an analytic solution of Eq. \eqref{kpb-eq}  of the form $\phi(\xi,\zeta,\tau)=\phi(\chi)$, where $\chi=K(\xi +\zeta) -\Omega\tau $ with $K$ and $\Omega$ denoting, respectively, the nondimensional constant   wave number (For simplicity,  we have considered $K_1$ and $K_2$, the constants along $\xi$ and $\zeta$ directions, as equal to $K$) and wave frequency.  Under this transformation Eq. \eqref{kpb-eq}    reduces   to an ordinary differential equation with respect to $\chi$, which can be solved by tanh method. There are several methods, e.g., inverse scattering method, using B{\"a}cklund transformation etc., however, the most convenient and efficient method is the tanh method \cite{tanh-method}.  By this method, a   solution of Eq. \eqref{kpb-eq} can be written as \cite{mas}
%\begin{widetext}
\begin{equation}
\phi(\chi)=\phi_0\left[2\left(1-\tanh~\chi\right) +\text{sech}^{2}~\chi\right], \label{shock-sol-kpb} 
\end{equation}
%\end{widetext}
where  $\phi_0=3C^2/25AB$,  $\Omega=D+(6/5)\eta_0K$,   and  $K=C/10B$. In particular, in absence of the weak transverse perturbation, i.e., perturbation along the $\zeta$ direction,  the KPB equation \eqref{kpb-eq} reduces to the Korteweg-de Vries Burger (KdVB) equation
\begin{equation}
\frac{\partial \phi }{\partial \tau }+A\phi 
\frac{\partial \phi }{\partial \xi }+B\frac{\partial ^{3}\phi }{\partial \xi
^{3}}=C\frac{\partial ^{2}\phi }{\partial \xi ^{2}}, \label{kdvb-eq} 
\end{equation}
 which has the following  shock solution (by tanh method)
 \begin{equation}
\phi(\chi^{\prime})=\phi_0\left[2\left(1-\tanh~\chi^{\prime}\right) +\text{sech}^{2}~\chi^{\prime}\right], \label{shock1-sol-kdvb} 
\end{equation}
 where $\chi^{\prime}=K^{\prime}\xi-\Omega^{\prime}\tau$ and  $\Omega^{\prime}=(6/5)\eta_0K^{\prime}$. That is, the wave speed of the KdVB shocks gets reduced, while the shock height remains unchanged.   The other form of   shock solution  of Eq. \eqref{kdvb-eq} can also be obtained as \cite{shock-kdvb}
 \begin{eqnarray}
&&\phi(\xi,\tau)=D_1-\frac{12C^2}{25AB\left(1+D_2e^{\psi}\right)^2},\notag\\ && \text{with}~\psi=-\frac{C}{5B}\xi+\left(\frac{ACD_1}{5B}-\frac{6C^3}{125AB^2} \right)\tau, \label{shock2-sol-kdvb}
\end{eqnarray}
where $D_1$ and $D_2$ are arbitrary constants. It can be shown that the profiles given by Eq. \eqref{shock2-sol-kdvb} will exhibit monotonic nature with increasing values of $\eta_0$ \cite{misra2012} or decreasing values of $H$. We also note that  the two shock (travelling wave) solutions \eqref{shock1-sol-kdvb} and \eqref{shock2-sol-kdvb} are obtained in two different methods, however, their   qualitative features will remain the same.    On the other hand, disregarding the     viscosity effect in Eq. \eqref{kpb-eq}, one can also recover the KP equation 
\begin{equation}
\frac{\partial }{\partial \xi }\left(\frac{\partial \phi }{\partial \tau }+A\phi 
\frac{\partial \phi }{\partial \xi }+B\frac{\partial^{3}\phi}{\partial \xi
^{3}}\right)+D\frac{\partial
^{2}\phi }{\partial \zeta ^{2}}=0, \label{kp-eq} 
\end{equation}
 which admits the following  soliton solution  (by the tanh method) \cite{tanh-method,mas} 
\begin{equation}
\phi(\xi,\zeta,\tau)=\frac{3(\widetilde{\Omega}-D)}{A}\text{sech}^{2}\left[\sqrt{\frac{\widetilde{\Omega}}{4B+D}}\left(\xi+\zeta-\widetilde{\Omega}\tau\right)\right], \label{soliton-sol-kp} 
\end{equation}
where $K=1$ and $\widetilde{\Omega}=4B+D$.  Furthermore, if we ignore the perturbation along $\zeta$ direction $(D=0)$ as well as the dissipation  $(C=0)$, Eq. \eqref{kpb-eq} reduces to the usual KdV equation with the following soliton solution
\begin{equation}
\phi(\xi,\tau)=\frac{3\widetilde{\Omega}}{A}\text{sech}^{2}\left[\sqrt{\frac{\widetilde{\Omega}}{4B}}\left(\xi-\widetilde{\Omega}\tau\right)\right]. \label{soliton-sol-kdv} 
\end{equation}
\subsection{Stability}
In order to study the stability of a travelling wave solution of  Eq. \eqref{kpb-eq} we apply the  technique  as in Ref. \citep{trav}. In the    frame $\chi=\xi+\zeta-\Omega\tau$ with $K=1$, Eq. \eqref{kpb-eq} reduces to an ordinary differential equation. The latter is then integrated twice   with respect to $\chi$, subject to the  
 boundary conditions, namely $\phi \rightarrow 0$, $d\phi/d\chi \rightarrow 0$, and $d^2 \phi/d\chi^2\rightarrow0$ as $|\chi|\rightarrow \infty$, to yield  
\begin{equation}
    B\frac{d^2 \phi}{d \chi^2}=C\frac{d \phi}{d
    \chi}-\frac{A}{2}\phi^2-(D-U)\phi. \label{stab}
\end{equation}
In the $(\phi, d\phi/d\chi )$ plane, Eq. (\ref{stab}) has two
singular points $(0,0)$  and $\left(2(U-D)/A, 0\right)$. While the former  corresponds to the   equilibrium downstream state,  the latter corresponds to the upstream one. Furthermore, the singular point $(0,0)$ is always a saddle point. The nature of the second one can be determined from the asymptotic behavior of the solution of the
form $\sim \exp(p \chi)$ \cite{karp} of the linearized form of  Eq.
(\ref{stab}) with $p$, given by
\begin{equation}
p=\frac{C}{2B}\left[1\pm
\sqrt{1-\frac{4B}{C^2}(\Omega-D)}\right].
\end{equation}
It follows that the singular point $\left(2(\Omega-D)/A, 0\right)$ is a stable focus or stable node according as $C^2 \lessgtr 4B(\Omega-D)$. While the stable focus always corresponds to the oscillatory nature,   the stable node corresponds to the monotonic nature of the solution.
\subsection{Numerical results }
 We  numerically analyze the exact analytic solutions of Eq. \eqref{kpb-eq}. We note that the parameters involved explicitly in the system are mainly $H$ and $\eta$.   They, respectively, appear in the coefficients of dispersion and dissipation. 
 We note that since $H\propto n_0^{-1/6}$, lower values of $H~(<1)$ corresponds to high-density plasma regimes. The values of $H>1$ is inadmissible, because otherwise the nonrelativistic $(V_{Fe}/c\ll1)$ quantum hydrodynamic model may not be valid \cite{hass2011}. 
   Furthermore, it follows that  quantum ion-acoustic shocks [Eqs. \eqref{shock-sol-kpb}, \eqref{shock1-sol-kdvb}]   can be either compressive or raraefactive according to when $H\lessgtr2/3$, 
   i.e., when one considers plasmas with the number density   satisfying $n_0\gtrless 1.3\times10^{31}$ cm$^{-3}$ and  the corresponding ion viscosity parameter satisfying  $\mu\gtrless0.0885\epsilon^{1/2}\eta_0~$kg/ms for some choice of values of $\epsilon$ and $\eta_0$ below the unity.  Typically, in the interiors of compact astrophysical objects such as  white dwarfs \cite{compact1,compact2,compact3}, the particle number density can vary in the range $10^{28}-10^{34}$ cm$^{-3}$ and the thermodynamic temperature $\lesssim10^7$ K.  Also, the  interiors of these stars may be considered as a plasma (e.g., carbon–oxygen white dwarf in a thermonuclear Supernova explosion) consisting of positively charged ions (nuclei) providing almost all the mass (inertia) and none of the pressure, as well as degenerate electrons providing all the pressure.  
   Now, in the  solutions \eqref{shock-sol-kpb} and \eqref{shock1-sol-kdvb},    the factor $C/10B$ mainly determines the steepness of the shocks. It is also clear that the nonlinear coefficient $A$ does not affect the shock steepness,  and the   dispersion coefficient $D$ affects neither the shock height nor its
steepness, it only plays a role in shifting the shock from its initial position with a passage of time. 
\begin{figure*}[ht]
\centering
\includegraphics[height=2.5in,width=6.0in]{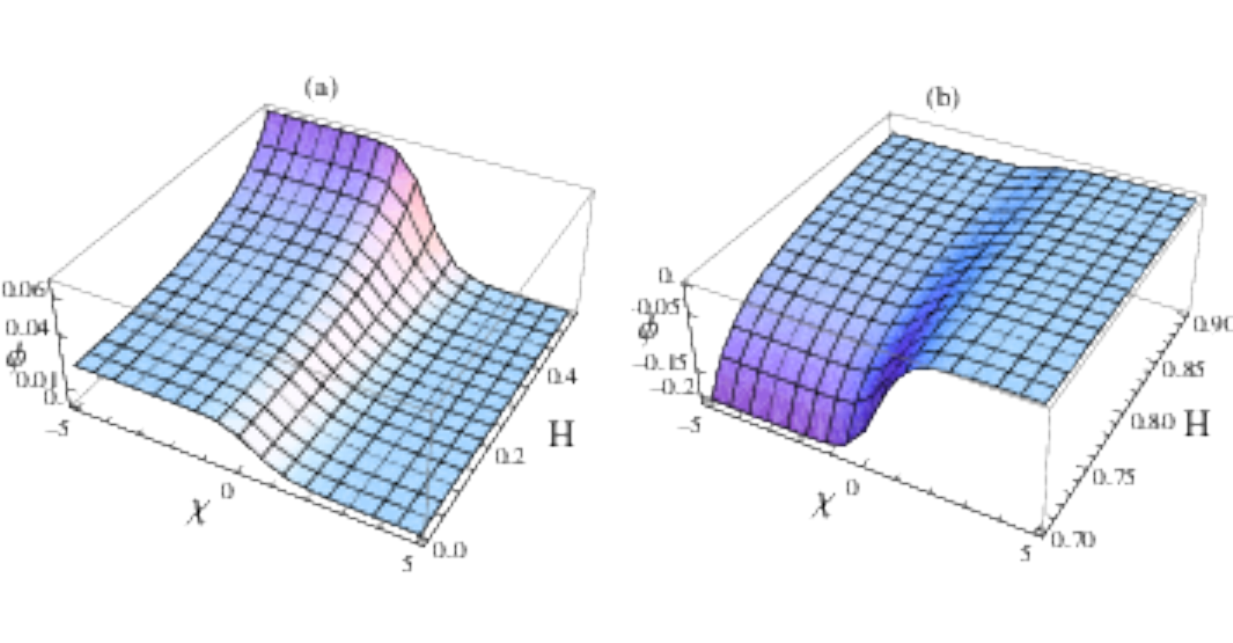}
\caption{Profiles of both the compressive [panel (a)] and rarefactive [panel (b)]  ion-acoustic shocks given by Eq. \eqref{shock-sol-kpb} are shown against $\chi$ and $H$ for a fixed $\eta_0=0.2$. The shock is  monotonic in nature.}
\label{fig:figure1}
\end{figure*}

\begin{figure}[ht]
\centering
\includegraphics[height=2.0in,width=3.0in]{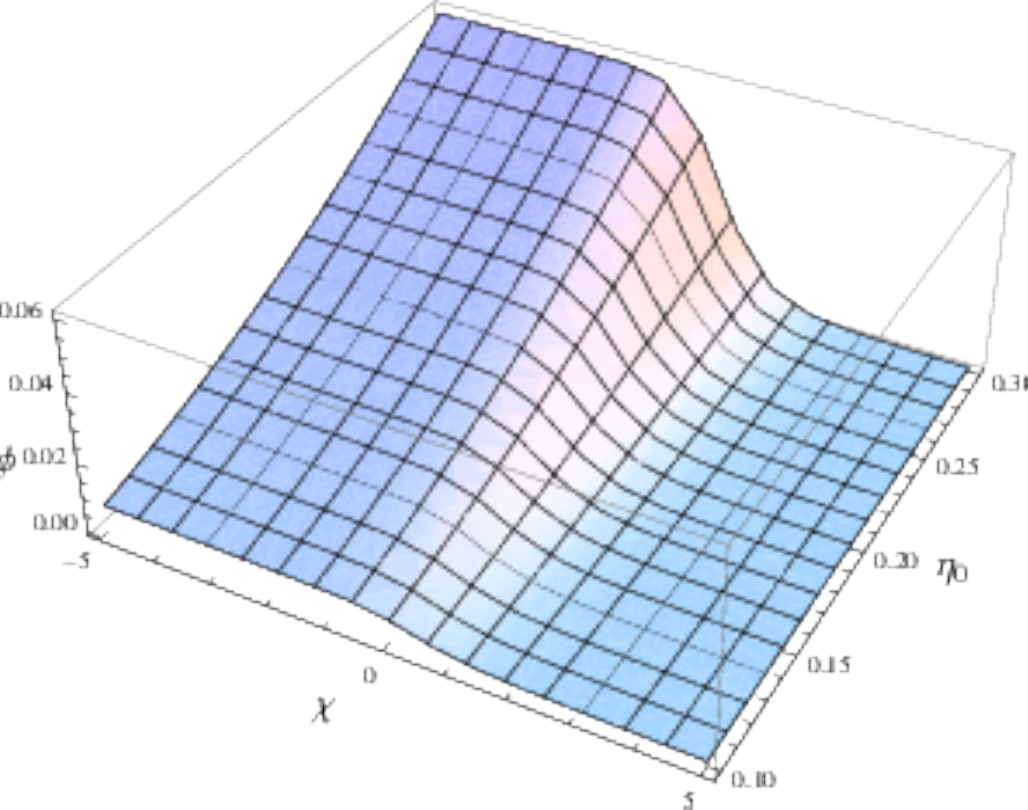}
\caption{Profile of the ion-acoustic shocks given by Eq. \eqref{shock-sol-kpb} are shown against $\chi$ and $\eta_0$ for a fixed $H=0.3$. The shock is of the compressive type and monotonic in nature.}
\label{fig:figure2}
\end{figure}

Typical    profiles of the shock solution \eqref{shock-sol-kpb} are shown in Figs. \ref{fig:figure1} and \ref{fig:figure2} with the variations of the parameters $H$ and $\eta$. Evidently, the compressive [Fig. \ref{fig:figure1}(a)]  and rarefactive shocks [Fig. \ref{fig:figure1}(b)] appear in the regimes $0<H<2/3$ and $2/3<H<1$ respectively. We here note that lower the values of $H$ in $0<H<1$, the higher is the particle number density.  Thus, in quantum plasmas, e.g., in the interior of white dwarfs, the formation of compressive QIA shocks is possible in the higher density regime with   $n_0\sim10^{31}-10^{34}$ cm$^{-3}$, whereas QIA shocks of the rarefactive type may occur in relatively a lower density regime with $n_0\sim10^{30}-10^{31}$ cm$^{-3}$. The corresponding   viscosity parameter  $\mu~\left(\equiv\epsilon^{1/2}\eta_0k_BT_{Fe}n_0/\omega_{pi}\right)$  range from   $0.003-12$ kg/ms for compressive shocks and $1.6\times10^{-4}-0.003$ kg/ms for rarefactive shocks (with $\epsilon=\eta_0=0.1$).    
It is also found that while the magnitude of the   height and strength increase for the  compressive shocks, the same decrease for the raraefactive ones  as the value of the quantum parameter $H$ gets increased. This is due to the fact that as $H$ increases in the regime $0<H<2/3$, the magnitude of the dispersive coefficient $B$ decreases [hence from Eq. \eqref{shock-sol-kpb} the magnitude of $\phi_0$ increases], whereas the same increases in $2/3<H<1$ [i.e., from Eq. \eqref{shock-sol-kpb} the magnitude of $\phi_0$ decreases].  In other words, for compressive shocks, as long as the ratio of electron plasmon energy  to  the Fermi thermal energy decreases and hence the corresponding dispersive effect is still weaker than the dissipation due to ion viscosity,  the  shock strength and height increase. However, for rarefactive shocks as $H$ increases, i.e., the dispersive effects become higher than the dissipation, the  magnitude of the  shock strength and height decrease. 
The effects of the ion  viscosity on the profiles of the compressive shocks (similar effects on the rarefactive shocks as well) are shown in Fig. \ref{fig:figure2}.
It is clear that for a fixed   $H=0.3$, i.e., $n_0\sim1.5\times10^{33}$ cm$^{-3}$ as the value of $\eta_0~(\lesssim1)$ increases from $0.1$ to $0.3$, i.e., $\mu$ increases from $0.73-2.2$ kg/ms (with $\epsilon=0.1$), the height and strength of the shock increase. Physically, increasing the ion viscosity  is equivalent to increasing the dissipation in the system, and  hence an increase in the shock strength. 

\begin{figure*}[ht]
\centering
\includegraphics[height=2.5in,width=6in]{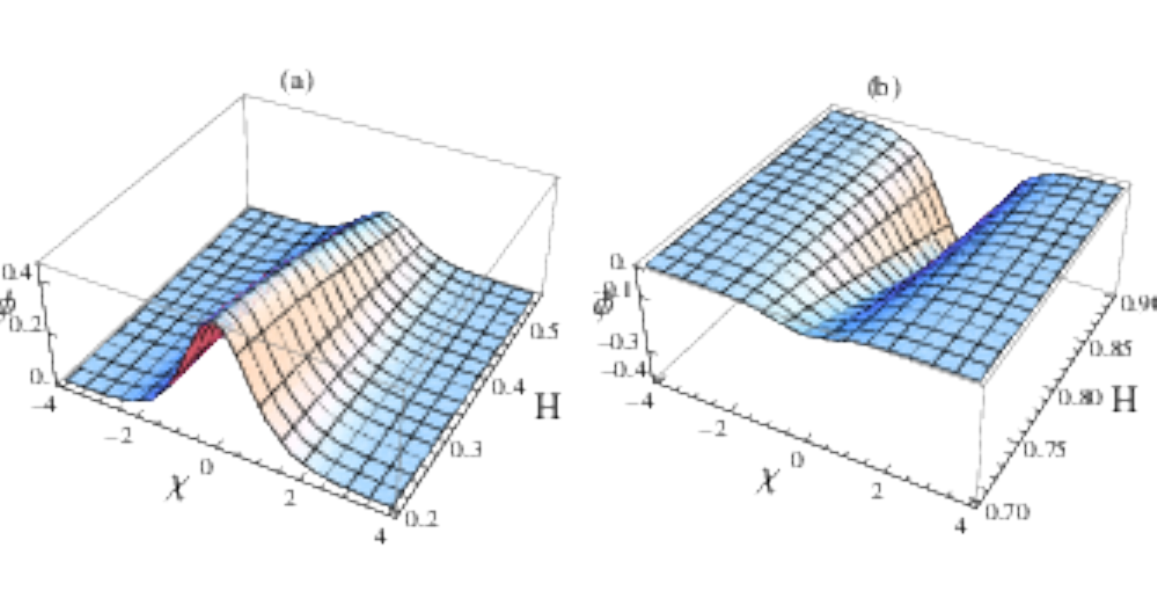}
\caption{Profiles of the ion-acoustic KP soliton given by Eq. \eqref{soliton-sol-kp} are shown against $\chi$ and the quantum parameter $H$.  While the left panel (a) shows that the compressive solitons form for $H<2/3$, the right panel (b) exhibits the existence of  raraefactive solitons in $H>2/3$. }
\label{fig:figure3}
\end{figure*}

Figure \ref{fig:figure3} exhibits the effects of the quantum parameter $H$ on the profiles of both compressive [Subplot (a)] and rarefactive [Subplot (b)] KP solitons [Eq. \eqref{soliton-sol-kp}]. It is found that   while the
amplitude of the compressive soliton decreases with an increase in $H$, the absolute value of the same for the rarefactive soliton gets  enhanced with   increasing values of $H$.

\begin{figure*}[ht]
\centering
\includegraphics[height=2.5in,width=6in]{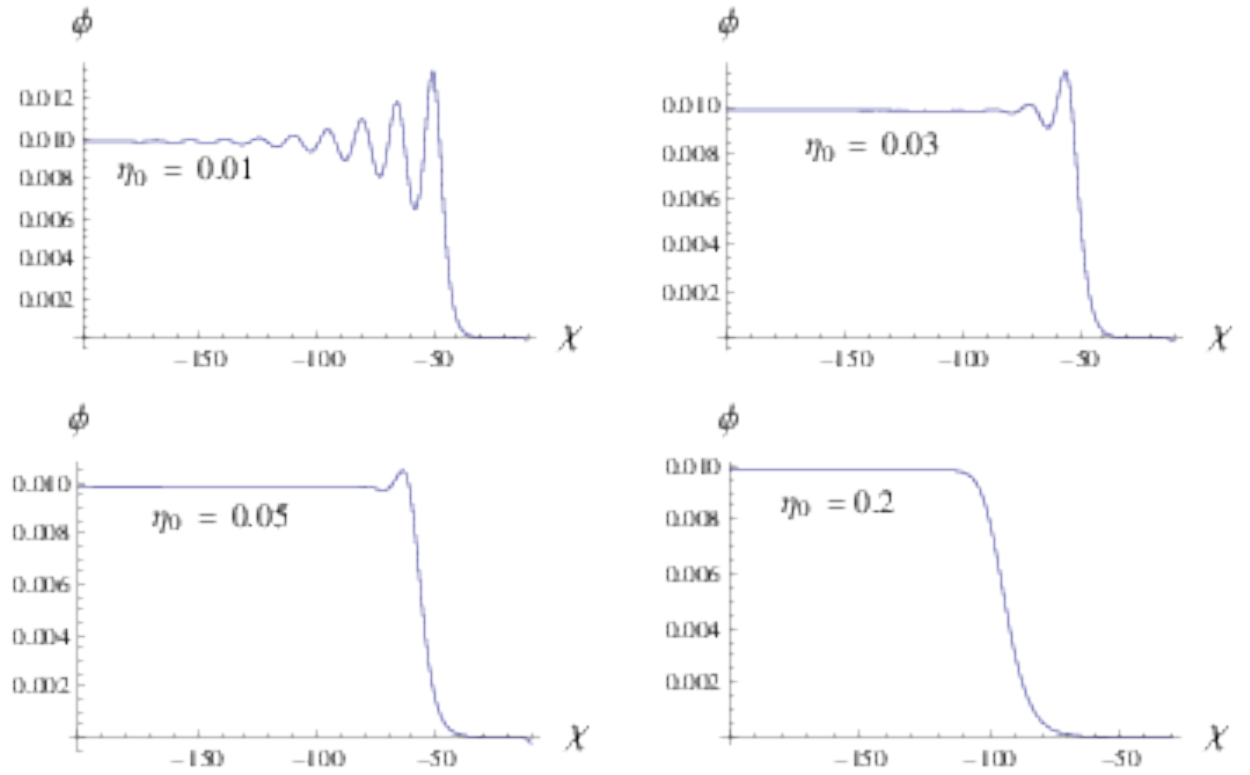}
\caption{Numerical solution of Eq. \eqref{kpb-eq} is plotted against $\chi$ to show the profiles of oscillatory and monotonic shocks for different values of $\eta_0$ (with a fixed $H=0.4$) as in the figure. A transition from oscillatory to monotonic   shocks is seen to occur at higher values of $\eta_0\lesssim1$.}
\label{fig:figure4}
\end{figure*}

 A numerical shock solution of Eq. \eqref{kpb-eq} is also presented in Fig. \ref{fig:figure4}  for a fixed $H=0.4$ (i.e., $n_0\sim2.7\times10^{32}$ cm$^{-3}$) and for different values of $\eta_0=0.01,~0.03,~0.05$ and $0.2$, which, respectively, correspond to the values $\mu=0.0097$ kg/ms, $0.029$ kg/ms, $0.0486$ kg/ms, and $0.19$ kg/ms for a scaling parameter $\epsilon=0.1$, i.e., almost the same values of $\eta_0$.  It basically exhibits how a shock profile transits from oscillatory to monotonic  one with increasing values of the viscosity parameter $\mu$.   We find that when $\eta_0$ or $\mu$ is very small, the shock wave admits an oscillatory profile. In this case, a train of solitons behind the shock is seen to form  whenever the dispersion   dominates over the dissipation (See the profile at $\eta_0=0.01$). However, when the dissipation overwhelms the dispersion and when the dissipative effect ($\mu$) is in nice balance with the nonlinearity arising from the nonlinear mode coupling of finite
amplitude QIA waves, the shock wave will have a monotonic behavior (See the profile at $\eta_0=0.2$).
%%%%%%%%%%%%%%%%%
%%%%%%%%%%%%%%%%%%%%%%%%%%%%%%%%%%%%%%%%%%%%%%%%%%%%%%%%%%%%%%%%%%%%%%%%%
\section{Arbitrary amplitude shocks}
So far we have investigated the evolution and properties of small-amplitude QIA solitary waves and shocks in quantum plasmas. However,  when the amplitude of the wave becomes larger the perturbation technique may no longer be valid, and one has to rely on some other method (See, e.g. Ref. \citep{misra2009}). So, for the propagation of  large-amplitude  QIA shocks  we transform the basic normalized equations \eqref{cont-eq-ni}-\eqref{moment-electron}  into a   moving frame of reference $\xi=l_x x+l_y y-Mt$, where $l_x^2+l_y^2=1$ and $M$ is the nonlinear wave speed (Mach number). Thus, from  Eqs. \eqref{cont-eq-ni}-\eqref{moment-electron}  we obtain
\begin{equation}
    -M\frac{dn_i}{d\xi}+l_x\frac{d}{d\xi}(n_i u)+l_y\frac{d}{d\xi}(n_i
    v)=0, \label{arbcon}
\end{equation}
\begin{equation}
    -M\frac{du}{d\xi}+\left(u l_x\frac{d}{d\xi}+v
    l_y\frac{d}{d\xi}\right)u=-l_x\frac{d\phi}{d\xi}+\eta\frac{d^2u}{d\xi^2}, \label{arbmomen1}
\end{equation}
\begin{equation}
    -M\frac{dv}{d\xi}+\left(u l_x\frac{d}{d\xi}+v
    l_y\frac{d}{d\xi}\right)v=-l_y\frac{d\phi}{d\xi}+\eta\frac{d^2v}{d\xi^2}, \label{arbmomen2}
\end{equation}
\begin{equation}
    \phi+\frac{1}{2}\left(1-n_e^{2/3}\right)+\frac{H^2}{2\sqrt{n_e}}\frac{\partial^2}{\partial
    \xi^2}\sqrt{n_e}=0, \label{arbquan1}
\end{equation}
Now,   using the
quasineutrality condition $n_e=n_i$, valid for long-wavelength limits (i.e., when the   length scale of excitation is much larger than the Fermi Debye length), we obtain from Eqs. (\ref{arbcon})-(\ref{arbquan1}) the following  nonlinear ordinary differential equation in $n_e$ as
\begin{eqnarray}
   &&\frac{H^2}{4}\left[\frac{d^2 n_e}{\partial
    \xi^2}-\frac{1}{2}\left(\frac{d n_e}{d\xi}\right)^2\right]+M \eta \frac{d n_e}{d\xi}\notag\\
    &&-\frac{M^2}{2}\left(1-n_e^2\right)+\frac{1}{2}n_e^2\left(1-n_e^{2/3}\right)=0, \label{singledif}
\end{eqnarray}
where we have imposed the boundary conditions $n_e\rightarrow 1$,
$\phi \rightarrow 0$, ${d n_e}/{d\xi} \rightarrow 0$ as
$|\xi|\rightarrow \infty$.

\begin{figure*}[ht]
\centering
\includegraphics[height=2.5in,width=6in]{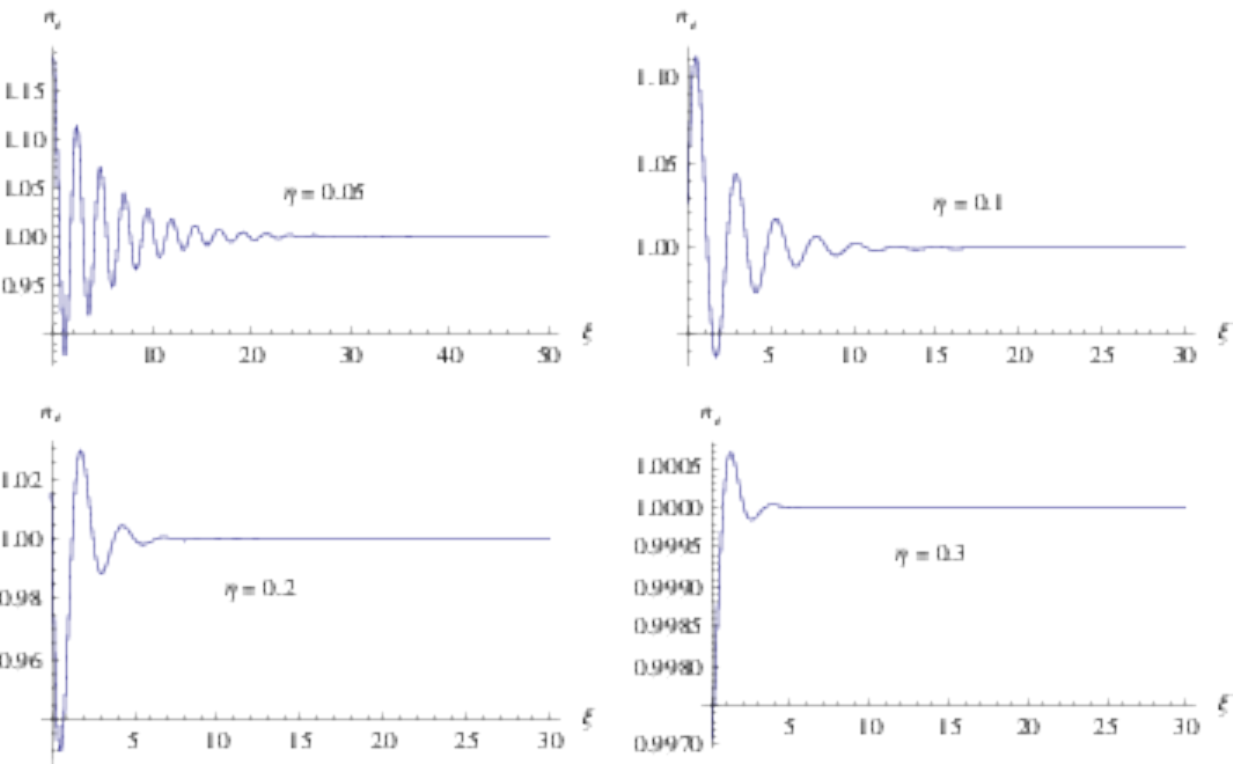}
\caption{Numerical solution of Eq. \eqref{singledif} is plotted against $\xi$ to show the profiles of oscillatory and monotonic shocks for different values of $\eta$ (with a fixed $H=0.8$ and $M=1.2$) as in the figure. A transition from oscillatory to monotonic   shocks is seen to occur at higher values of $\eta\lesssim1$.}
\label{fig:figure5}
\end{figure*}
%%%%%
\begin{figure*}[ht]
\centering
\includegraphics[height=2.5in,width=6in]{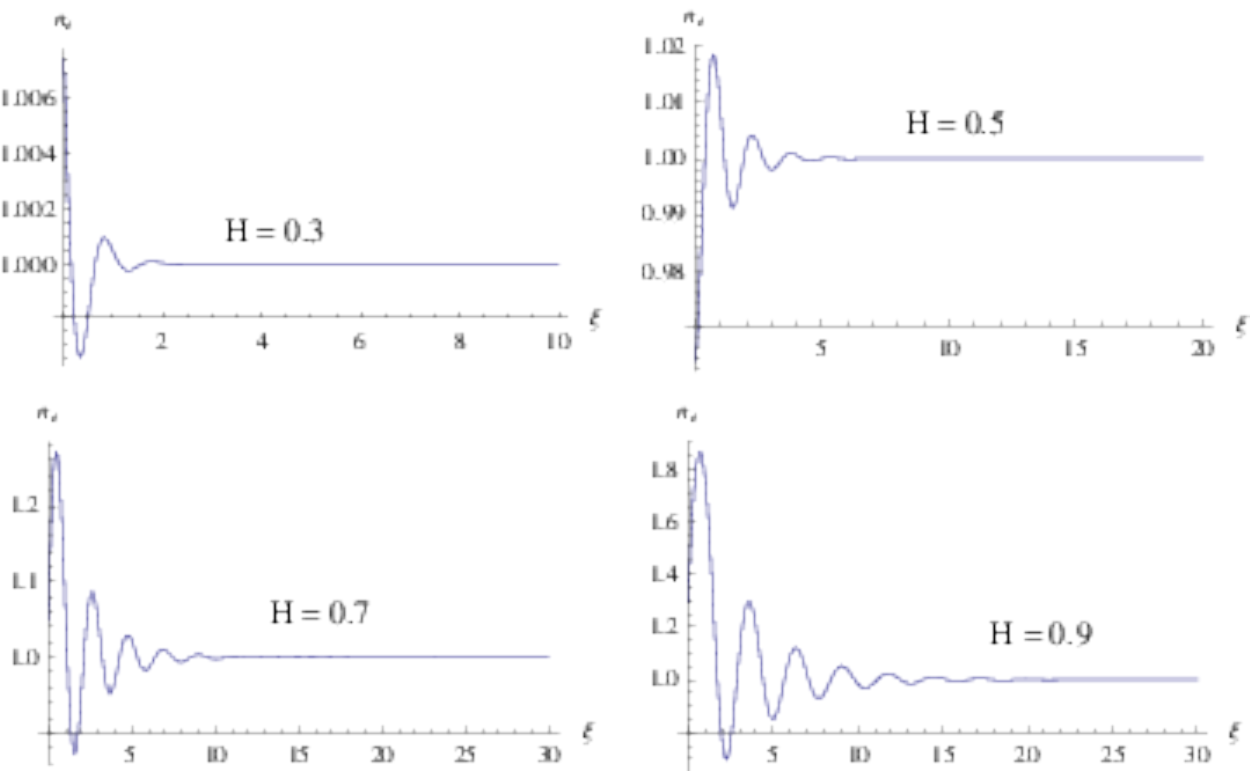}
\caption{Numerical solution of Eq. \eqref{singledif} is plotted against $\xi$ to show the profiles of oscillatory and monotonic shocks for different values of $H$ (with a fixed $\eta=0.1$ and $M=1.2$) as in the figure. A transition from monotonic to oscillatory shocks is seen to occur at higher values of $H<1$.}
\label{fig:figure6}
\end{figure*}
 In order to study the influence of the Bohm potential  as well as the damping parameter ($\eta$)  on the large amplitude QIA shocks, we numerically solve Eq. \eqref{singledif}. To this end we use   MATHEMATICA and apply the finite difference scheme.  The results are displayed in Figs. \ref{fig:figure5} and \ref{fig:figure6} for different values of $\eta$ and $H$ but with a fixed $M=1.2$. From Fig. \ref{fig:figure5}  it is found that when the dissipative effect is weaker, a train of solitons  is formed in front of the shock, resulting into the formation of oscillatory shocks (See the profile at $\eta=0.05$). As the value of $\eta$ starts increasing, the number of solitons gets reduced, and eventually at a large value of $\eta$, a transition to monotonic shocks occurs (See the profile at $\eta=0.3$). On the other hand, Fig. \ref{fig:figure6} shows that  as the dispersive effect due to quantum fluctuations becomes stronger, i.e., as the  value of $H$ increases [from $0.3$ to $0.9$, i.e.,  one approaches from high-density ($n_0\sim10^{33}$ cm$^{-3}$) to  low-density ($n_0\sim10^{30}$ cm$^{-3}$)  regimes with decreasing values of the ion viscosity parameter $\mu=2.3-0.001$ kg/ms  with a fixed $\eta=0.1$], the monotonic shock profile transits into an oscillatory one.  It is observed that the large amplitude QIA shocks preserve its monotonicity as long as the dissipation parameter is dominant over the dispersion one. We also mention that keeping the values of $H$ and $\eta_0$   fixed, say at $H=0.8$ and $\eta=0.1$ (as in Fig. \ref{fig:figure5}), as $M$ increases from $M=1.2$, the number of oscillations behind the shock decreases, and a transition from oscillatory to monotonic shock occurs at $M=2.8$. This is expected as in Eq. \eqref{singledif}, $M$ appears not only in the coefficients of nonlinear terms, but also in the dissipative term $(\propto\eta)$. Thus, for   the large-amplitude QIA shocks, a upper limit of $M$ also exists below which the shock profile remains oscillatory in nature.   
\section{Conclusion}
In this paper, we have studied the nonlinear propagation of ion-acoustic solitary waves and shocks  in an electron-ion quantum plasma with  the effects of  ion kinematic viscosity as well as the density correlation due to quantum fluctuation (quantum force associated with the Bohm potential). While the former plays a dissipative role, and thus favors the formation of shcoks,   the latter  provides the dispersion, thereby favoring the formation of  ion-acoustic solitons in quantum plasmas.      Both the small- and large-amplitude perturbations of QIA waves are considered. It is shown that in the small-amplitude limit, the propagation of QIA waves can be described by a KPB-type equation which admits  compressive or rarefactive solitay waves and shocks according to when the quantum parameter $H<2/3$ or $>2/3$  i.e., in quantum plasmas with  number density   satisfying $n_0\gtrless 1.3\times10^{31}$ cm$^{-3}$ and  the ion viscosity parameter satisfying  $\mu\gtrless0.09\epsilon^{1/2}\eta_0~$kg/ms for some choice of values of $\epsilon$ and $\eta_0$ lower than unity.  Such parameter regimes can be achieved, e.g., in the interiors of compact astrophysical objects like  white dwarfs where the particle number density can vary in the range $10^{28}-10^{34}$ cm$^{-3}$ and the thermodynamic temperature $\lesssim10^7$ K.  A numerical shock solution of the KPB equation is also obtained to show how an oscillatory shock profile transits into a monotonic one and vice versa.

 On the other hand, the properties of large-amplitude QIA shocks are studied in a frame of reference moving with a constant speed $M$. For simplicity, we have considered the case when the dispersion due to charge separation of particles is negligible. This approximation is valid for long-wave length ion pulses or when the system  length scale of excitation is much larger than the Fermi Debye length.    It is shown that the evolution of large-amplitude QIA shocks can be governed by a nonlinear ordinary differential equation which can not be reduced to a well-known energy-like equation for a pseudoparticle with a pseudopotential. However, the equation is solved numerically, and it is shown that the large-amplitude shocks exist only of the compressive type.  It is also found that a transition from oscillatory to monotonic shocks occurs not only with increasing values of $\eta$ or $\mu$, but also with some finite $M~(>1)$    depending on  the consideration of other parameter values of  $H$ and $\eta$.
 
 To conclude, the results should be useful for the nonlinear propagation of ion-acoustic solitary waves and shocks in degenerate plasmas such as those in compact astrophysical objects, e.g.,  white dwarfs.  The results may be of   importance for the excitation of quantum ion-acoustic (electrostatic) perturbations in   laboratory as well as for laser produced plasmas. 
%%%%%%%%%%%%%%%%%%%%%
\section*{Acknowledgement}
{Authors gratefully acknowledge useful comments from the anonymous referees which improved the manuscript in its present form.  This research was partially supported by   the SAP-DRS (Phase-II), UGC, New Delhi, through sanction letter No. F.510/4/DRS/2009 (SAP-I) dated 13 Oct., 2009, and by the Visva-Bharati University, Santiniketan-731 235, through Memo No.  REG/Notice/156 dated January 7, 2014.}
%%%%%%%%%%%%%%%%%%%%%%%%%

\end{document}